\documentclass[twoside,journal,10pt]{IEEEtran}
\usepackage{makecell}
\usepackage{hyperref}
\usepackage{array}
\usepackage{graphicx,amssymb,amsmath}
\usepackage{multicol}
\usepackage[noadjust]{cite}
\usepackage{setspace}
\usepackage{subfigure}
\usepackage{float}
\usepackage{listings}
\usepackage {url}
\usepackage{stfloats}
\usepackage{amsthm,pifont}
\usepackage{flushend}
\usepackage{cases,subeqnarray}
\usepackage{bm,multirow,bigstrut}
\usepackage{textcomp}
\usepackage{latexsym,bm}
\usepackage{booktabs}
\usepackage{xcolor}
\usepackage{mathtools}
\usepackage{dsfont}
\usepackage{extarrows}
\usepackage{epsfig}
\usepackage{epsfig}
\usepackage{epstopdf}
\usepackage{colortbl}
\usepackage{algpseudocode}
\usepackage{algorithmicx,algorithm}
\lstdefinelanguage{json}{
  basicstyle=\ttfamily\footnotesize,
  numbers=left,
  numberstyle=\tiny\color{gray},
  stepnumber=1,
  numbersep=6pt,
  showstringspaces=false,
  breaklines=true,
  frame=single,
  backgroundcolor=\color{gray!5},
  literate=
   *{0}{{{\color{blue}0}}}{1}
    {1}{{{\color{blue}1}}}{1}
    {2}{{{\color{blue}2}}}{1}
    {3}{{{\color{blue}3}}}{1}
    {4}{{{\color{blue}4}}}{1}
    {5}{{{\color{blue}5}}}{1}
    {6}{{{\color{blue}6}}}{1}
    {7}{{{\color{blue}7}}}{1}
    {8}{{{\color{blue}8}}}{1}
    {9}{{{\color{blue}9}}}{1}
}
\theoremstyle{plain}

\theoremstyle{plain}

\usepackage{amsmath}

\definecolor{Gray}{gray}{0.85}
\definecolor{lightgray}{RGB}{240,240,240}
\definecolor{lightblue}{RGB}{220,235,255}
\definecolor{lightgreen}{RGB}{230,250,230}
\IEEEoverridecommandlockouts
\begin{document}

\date{}


\title{\Large \bf NetMCP: Network-Aware Model Context Protocol Platform for LLM Capability Extension\\
}



\author{
Enhan~Li, Hongyang~Du\textsuperscript{*}, Kaibin~Huang \\
The University of Hong Kong \\
\textsuperscript{*}Corresponding author
}


\maketitle

\begin{abstract}
Large Language Models (LLMs) remain static in functionality after training, and extending their capabilities requires integration with external data, computation, and services.
The Model Context Protocol (MCP) has emerged as a standard interface for such extensions, but current implementations rely solely on semantic matching between users' requests and server function descriptions, which makes current deployments and simulation testbeds fragile under latency fluctuations or server failures.
We address this gap by 
enhancing MCP tool routing algorithms with real-time awareness of network and server status. 
To provide a controlled test environment for development and evaluation, we construct a heterogeneous experimental platform, namely Network-aware MCP (NetMCP), which offers five representative network states and build a benchmark for latency sequence generation and MCP server datasets.
On top of NetMCP platform, we analyze latency sequences and propose a Semantic-Oriented and Network-Aware Routing (SONAR) algorithm, which jointly optimizes semantic similarity and network Quality of Service (QoS) metrics for adaptive tool routing. 
Results show that SONAR consistently improves task success rate and reduces completion time and failure number compared with semantic-only, LLM-based baselines, demonstrating the value of network-aware design for production-scale LLM systems. The code for NetMCP is available at https://github.com/NICE-HKU/NetMCP.


\end{abstract}

\section{Introduction}
\begin{figure}[!t]
\centering
\includegraphics[width=3 in]{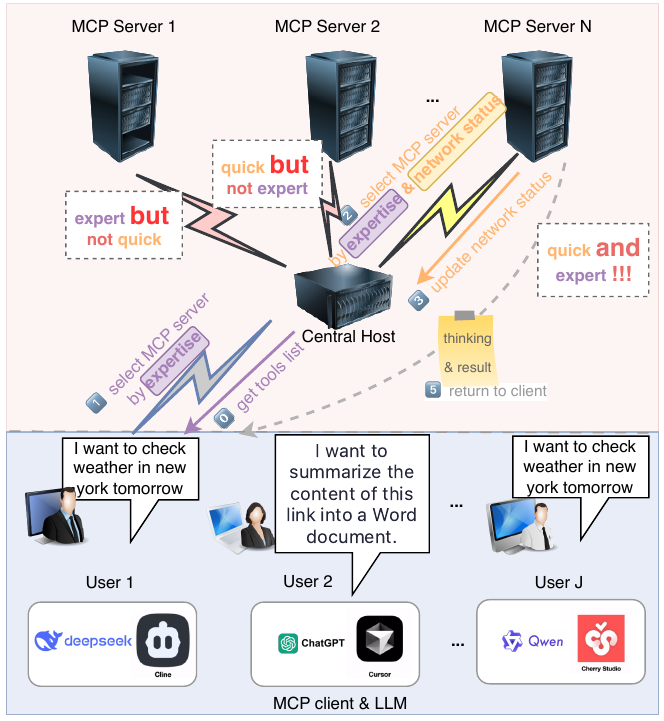}
\caption{Motivation for considering network states and designing a joint optimization algorithm for MCP tool routing. Naive routing policies based only on tool expertise, risk delays or failures under poor network conditions, while policies based only on network conditions may select irrelevant tools. 
Our algorithm resolves this trade-off to find the optimal tool that is both highly expert and network-efficient.}
\label{fig_1}
\end{figure}
Large Language Models (LLMs) have advanced rapidly in recent years and are now deployed in a wide range of applications such as search engines, productivity software, and research tools~\cite{achiam2023gpt4, grattafiori2024llama3, yang2025qwen3}. 
Their ability to generate fluent text, reason across domains, and assist human decision-making has positioned them as a central component of modern artificial intelligence. 
With broader adoption, expectations for LLMs have also expanded. Users increasingly require access to timely knowledge, reliable reasoning across specialized domains, and support for complex computational tasks. 
However, LLMs in their pure form are static models, with knowledge and capabilities fixed in their parameters once the training process is complete. 
Without acquiring new information or connecting to external environments, they face clear limitations when applied to real-world tasks.
As a result, the expansion of LLM capabilities increasingly depends on access to external tools typically hosted on distributed servers, such as web search for real-time information, location services for context awareness, and domain-specific systems for specialized functions \cite{Toolcallsurvey1,Toolcallsurvey2,Toolcallsurvey3,toolformer}.

However, connecting LLMs with external resources, whether tools, services, or other LLMs, has historically been complex. Integrations often required custom interfaces or task-specific plugins, which led to fragmented solutions that were difficult to maintain and scale. To address this issue, several interaction protocols have been proposed~\cite{protocol_survey}, including the Model Context Protocol (MCP)~\cite{MCP, mcp_introduction, singh2025survey_mcp}, Agent Network Protocol (ANP)~\cite{anp_github,anp_website}, Agent Communication Protocol (ACP)~\cite{ACP, ibm_beeai_acp}, and Agent-to-Agent (A2A)~\cite{google_a2a} protocols. 
Among these efforts, MCP has developed into a broadly adopted and open standard, providing the most complete infrastructure for integrating LLMs and diverse data sources and tools~\cite{MCP}. Through MCP, LLMs can route requests to external tools when handling tasks such as real-time information retrieval, complex computation, and system control~\cite{singh2025survey_mcp}.

A central challenge for MCP and related protocols is how available tools are presented and selected, which could be considered a fundamental problem for LLMs' interfaces with external resources.
As the number of tools continues to grow, exposing all tool descriptions to the model is neither efficient nor scalable, and existing selection algorithms that focus solely on this issue overlook the complexity of real deployment environments where network conditions play a decisive role. At its core, the challenge is to identify the right tools for a given user query in real LLM-environment interactions. However, most approaches \cite{MCPZero, RAGMCP} reduce this to semantic similarity matching between user queries and tool descriptions by employing Retrieval-Augmented Generation (RAG) to perform tool selection.
Taking MCP Zero \cite{MCPZero} as an example, it employs a two-stage strategy. By considering that each server hosts a different set of tools, the first stage selects a subset of top-K MCP servers by measuring semantic similarity between the user query and server descriptions. The second stage then refines the choice by matching the query against individual tool descriptions and selecting the final top-K tools. While this improves relevance at the semantic level, the approach remains an {\textit{offline}} matching process that ignores the {\textit{online}} network status of the MCP servers. As shown in Fig.~\ref{fig_1}, this limitation can lead to mismatches where the tool that appears most relevant semantically is hosted on a server that is unavailable, overloaded, or subject to high latency. In such cases, the task execution may be less effective than selecting a slightly less relevant but more stable tool.

Therefore, a robust tool routing algorithm suitable for production environments must incorporate real-time awareness of network status, such as latency, availability, and load, in addition to semantic relevance.
However, existing approaches~\cite{MCPZero,RAGMCP} are limited to semantic matching due to the absence of experimental platforms that systematically incorporate network monitoring. 
Existing benchmarks \cite{mcpbench,mcpradar,mcpuniverse,livemcpbench,liu2025mcpeval} differ in scope and tool coverage, but all neglect real network conditions when assessing tool efficiency and task accuracy. This gap highlights two central challenges:
\begin{itemize}[noitemsep, topsep=0pt]
\item {\textbf{C1.}} How to provide an experimental environment that integrates heterogeneous network states to support systematic evaluation of MCP-based tool selection under various conditions.
\item {\textbf{C2.}} How to consider network-aware metrics to achieve robust tool selection in dynamic environments.
\end{itemize}

To tackle {\textbf{C1}}, our idea is to build an experimental platform that combines MCP servers with configurable network environments. Such a platform should provide flexible server datasets and allow systematic injection of diverse network states, making it possible to reproduce the dynamics that real LLM-tool systems face. This foundation allows algorithms to be tested and compared under realistic conditions rather than in purely semantic benchmarks.
Addressing the {\textbf{C2}} requires rethinking the tool routing algorithm. Instead of treating it as a purely semantic retrieval task, we frame it as a joint optimization problem that balances tool expertise with network health. Semantic similarity ensures that a tool is functionally relevant, while latency and stability determine whether it can deliver results in practice. A selection mechanism that unifies these two dimensions can thus avoid the mismatch between semantic optimality and deployment feasibility.
In summary, this paper makes the following contributions:

\begin{itemize}
\item {\textbf{Environment:}} We construct the Network-aware MCP (NetMCP) platform, a heterogeneous experimental environment that emulates real-world networks, incorporating five typical network states (e.g., fluctuating latency, intermittent outages, high latency, high jitter, and near-ideal conditions), to comprehensively evaluate the robustness of tool routing strategies under various network conditions, providing a reliable foundation for algorithm design and validation.


\item {\textbf{Algorithm:}} We propose the Semantic-Oriented and Network-Aware Routing (SONAR) algorithm. This algorithm continuously monitors the health status of MCP servers by leveraging historical latency sequences from each request, and incorporates a joint optimization algorithm that performs weighted fusion of network Quality of Service (QoS) metrics and semantic similarity, enabling intelligent and adaptive tool selection.

\item {\textbf{Experiment:}} We establish a comprehensive experimental benchmark incorporating automated latency sequence generation and MCP server dataset construction. On this benchmark, we compare the proposed method against multiple baseline approaches, including pure semantic matching, and LLM-based matching. Experimental results demonstrate that our algorithm significantly outperforms existing solutions in task success rate and average completion time, which validates both its effectiveness and practical superiority.
\end{itemize}

The paper is structured as follows. Section~\ref{sec:relatedwork} reviews background on tool-augmented LLMs and the MCP, and highlights the absence of network awareness in current tool selection methods. Section~\ref{subsec:experimental_platform} then introduces the design of our NetMCP platform for network state monitoring and benchmarking, and Section~\ref{subsec:joint_optimization} presents the proposed joint tool routing algorithm SONAR. Section~\ref{experiment} reports the experimental results followed by discussions of future work in Section~\ref{sec:future_work} and conclusions presented in Section~\ref{conclusion}.

\section{Related Work}
\label{sec:relatedwork}
\subsection{Tool-Augmented LLMs}
LLMs~\cite{achiam2023gpt4,liu2024deepseek,yang2025qwen3} demonstrate strong reasoning and generation capabilities, yet their knowledge remains bounded by the parameters fixed after training. To overcome this limit, LLMs are increasingly augmented with external tools that provide access to real-time information and domain-specific services, such as APIs, calculators, and Python interpreters~\cite{Toolcallsurvey1,Toolcallsurvey2,Toolcallsurvey3}. This integration is critical for moving beyond static text generation toward solving complex real-world tasks.
Early work explored narrow integrations with task-specific tools, such as search engines or symbolic solvers \cite{nakano2021webgpt,karpas2022mrkl}. Subsequent frameworks introduced more general protocols for reasoning and multi-tool usage, notably ReAct \cite{react} and AutoGen \cite{wu2024autogen}, which coupled tool invocation with step-by-step reasoning. In parallel, training-based approaches such as Toolformer \cite{toolformer}, ToolLLM \cite{qin2023toolllm}, and Tool-MVR \cite{ToolMVR} taught models to generate tool calls through self-supervised or instruction-tuned learning autonomously.
Another line of work has emphasized context-based methods, exemplified by ChatGPT Function Calling \cite{functioncall} and HuggingGPT \cite{shen2023hugginggpt}, which enable tool usage without retraining by injecting function definitions directly into the prompt.
These advances highlight the central role of tool augmentation in extending LLM functionality. Beyond enhancing accuracy and efficiency, tool use provides a practical path for LLMs to access external knowledge, interact with dynamic environments, and execute actions that go beyond text prediction. Recognizing this importance has also motivated the development of protocols such as the MCP \cite{mcp_introduction}, which aims to standardize tool integration, discovery, and coordination across AI platforms.

\subsection{Model Context Protocol}
The MCP~\cite{mcp_introduction} has emerged as a widely adopted standard for connecting LLMs with external tools through a JSON-RPC interface~\cite{MCP}. 
Inspired by the Language Server Protocol (LSP)~\cite{lsp}, MCP replaces fragmented, task-specific connectors with a universal protocol that supports dynamic tool discovery, orchestration, and human-in-the-loop oversight~\cite{singh2025survey_mcp,sarkar2025survey}. Its design allows LLMs to interact with diverse services in a consistent manner, lowering the cost of integration and enabling large-scale interoperability across platforms.
Since its release, MCP has been integrated into major AI ecosystems~\cite{achiam2023gpt4,yang2025qwen3} and has led to the development of servers across domains such as file systems~\cite{filemcp}, databases~\cite{databasemcp}, and web services~\cite{searchmcp}. 
Building on this growing ecosystem, new evaluation and benchmarking frameworks have been proposed, including MCPBench~\cite{mcpbench}, MCP-Radar~\cite{mcpradar}, MCP-Universe~\cite{mcpuniverse}, LiveMCPBench~\cite{livemcpbench}, MCPEval~\cite{liu2025mcpeval}, which measure efficiency, accuracy, and scalability of MCP-based interactions.
Beyond benchmarking, MCP-Zero \cite{MCPZero} introduces autonomous tool discovery and scalable capability extension, while AgentMaster \cite{liao2025agentmaster} integrates MCP with A2A \cite{google_a2a} in a modular multi-agent framework for coordinated multimodal AI interaction.

These developments confirm MCP's role as the core infrastructure for tool-augmented LLMs. However, existing work largely evaluates tool usage regarding accuracy, semantic relevance, and throughput, while paying little attention to network-level conditions such as latency, stability, and downtime. In practice, these factors directly influence the success and responsiveness of LLM–tool interaction. Addressing this gap requires extending MCP with mechanisms to monitor network state and incorporating these metrics into tool selection.
Our Network-Aware MCP framework addresses this gap by incorporating network state into MCP and considering latency as a critical factor for real-world performance.


\section{NetMCP Platform Design}
\label{subsec:experimental_platform}

This section describes the architecture and workflow of the NetMCP platform, 
an experimental environment that evaluates LLM extensions via MCP while explicitly accounting for network effects. The platform enables testing of existing tool routing methods and supports the development of new network-aware algorithms.

\begin{figure}[!t]
\centering
\includegraphics[width=3 in]{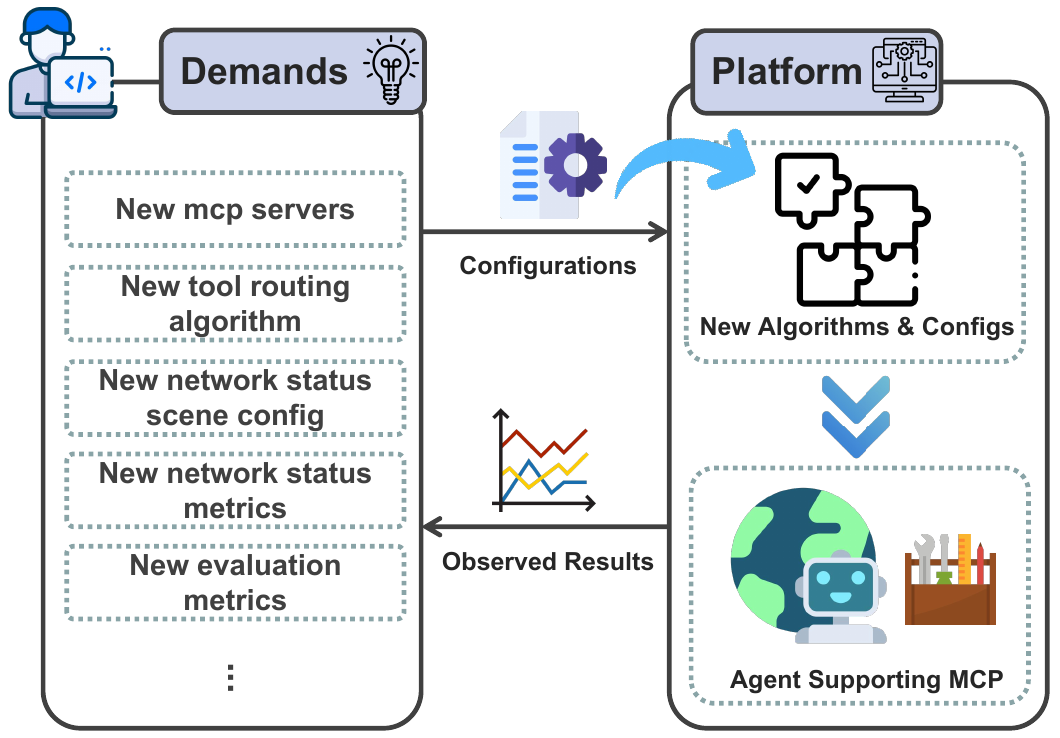}
\caption{The design of the NetMCP platform.}
\label{fig_2}
\end{figure}

\subsection{Platform Architecture}
\label{sec:Platform}

\begin{figure*}[!t]  
\centering
\includegraphics[width=0.9\textwidth]{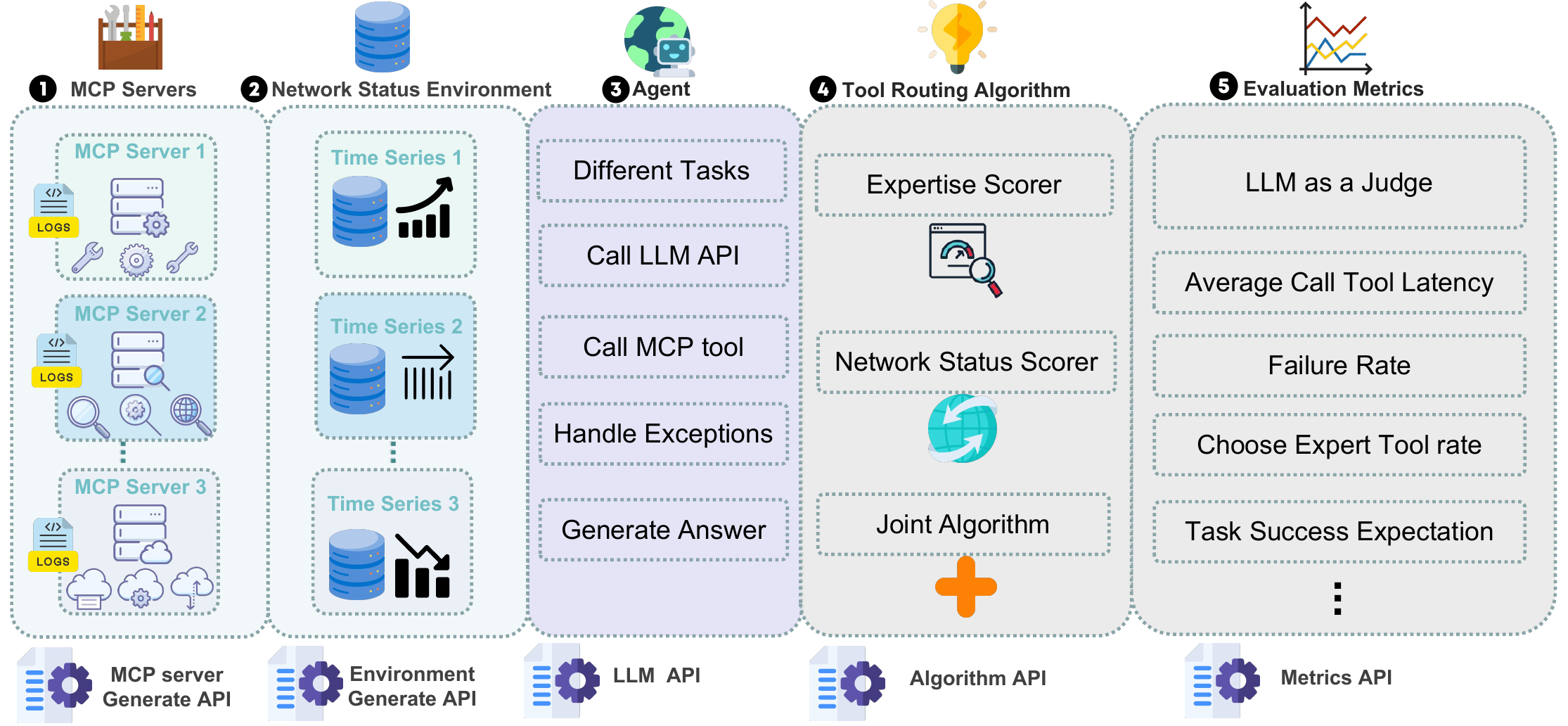}  
\caption{Modular architecture of the NetMCP platform. The NetMCP comprises five core modules, i.e., MCP servers, network status environment, agent, tool routing algorithm, and evaluation metrics, interacting through well-defined APIs to enable reproducible evaluation of network-aware tool routing algorithms under diverse and controllable conditions.}
\label{fig_3}
\end{figure*}

Although existing MCP benchmarks provide diverse scenarios and tasks, they remain limited in two critical aspects: 
\begin{itemize}
\item \textit{Limited scalability of validation.} They do not support convenient multi-scenario and large-scale testing of tool routing algorithms with virtual MCP servers. Without such a fast testing environment deployment, researchers cannot efficiently iterate or compare methods across diverse conditions in a reproducible way.
\item \textit{Lack of network realism.} They neglect latency, availability, and stability, reducing tool routing to a purely semantic matching task. However, these network factors could be decisive in real deployments, where the most semantically relevant tool may fail due to link congestion or downtime.
\end{itemize}

Therefore, an ideal experimental platform for MCP tool routing should enable the development and evaluation of realistic algorithms that remain effective across diverse and complex network environments, with minimal development overhead. 
It should provide configurable network conditions, scalable MCP server datasets, and APIs that make implementing and testing new routing algorithms straightforward. 
As illustrated in Figure~\ref{fig_2}, such a platform allows users to flexibly configure network settings and rapidly experiment with novel routing algorithms.

To achieve this, adopting a modular architecture that defines the critical components in the MCP-supported environment and agent is essential. With each, modules expose well-defined interfaces to avoid cross-layer dependencies. Fig.~\ref{fig_3} shows the overview of our proposed NetMCP platform's modular architecture, consisting of MCP servers, network status environment, agent supporting MCP, tool routing algorithm, and evaluate metrics. Each module is accessible through standardized APIs, enabling researchers to explore, extend, and integrate new mechanisms with minimal effort.
\begin{figure}[t]
\centering
\begin{lstlisting}[language=json, numbers=none]
{
  "last_time": "24h",
  "hybrid_scenario": {
    "High_Latency_Server": {
      "base_latency": "350ms",
      "std_dev": "20ms"
    },
    "Low_Latency_Server": {
      "base_latency": "30ms",
      "std_dev": "5ms"
    },
    "Intermittent_Outage_Server": {
      "base_latency": "30ms",
      "std_dev": "5ms",
      "failure_config": {
        "type": "intermittent",
        "probability": 0.5,
        "duration": ["30min", "100min"],
        "severity": ["1000ms", "1000ms"]
      }
    },
    "Fluctuate_Burst_Server": {
      "base_latency": "150ms",
      "std_dev": "20ms",
      "periodicity": {
        "amplitude": "200ms",
        "period": "360ms",
        "phase_shift": 0
      }
    },
    "High_Jitter_Server": {
      "base_latency": "100ms",
      "std_dev": "70ms"
    }
  }
}
\end{lstlisting}
\caption{Example of a hybrid scenario configuration generated by the NetMCP platform, simulating diverse network behaviors including high latency, low latency, intermittent outages, fluctuating bursts, and high jitter.}
\label{fig:hybrid_scenario}
\end{figure}

  

\textbf{Module 1: MCP Servers.}
This module enables the construction of a heterogeneous pool of external MCP tools that the LLM agent can invoke, serving as the foundation for scalable and reproducible experimentation and offering several core capabilities: 

\begin{itemize}
    \item {\textit{Rapid dataset generation from real-world APIs.}}
    This module enables keyword-driven retrieval and automatic generation of real-world MCP server datasets. Users can specify functional domains such as "websearch", "database", "AI coding", or "remote" to to fetch existing tools and organize them into domain-specific repositories. 
    This capability accelerates dataset curation while preserving the semantic diversity of real MCP services, and also establishes a scalable foundation for systematic benchmarking and extensible platform development.
    
    \item {\textit{Flexible simulation of large-scale server clusters.}}
    This module provides dynamic mocking functionality, allowing users to generate multiple virtual MCP servers from minimal template configurations. Developers can scale server populations on demand without deploying physical instances, making it ideal for large-scale ablation studies and rapid prototyping of routing algorithms. For instance, starting from a single real server such as Exa~\cite{exa}, which supports web search, users can automatically instantiate a cluster of 20 functionally similar virtual servers, each independently configurable with distinct network states for controlled evaluation of routing algorithm performance.

    \item {\textit{Dual-mode execution for both live benchmarking and controlled simulation.}}
    This module supports two complementary execution modes to accommodate different experimental needs. In live mode, tools are invoked on actual MCP servers\footnote{Examples of actual MCP servers:\\ 
    \: \url{https://smithery.ai/server/exa}\\ 
    \: \url{https://smithery.ai/server/@tdu-naifen/duckduckgomcp}\\
    \: \url{https://smithery.ai/server/brave}}, enabling end-to-end benchmarking under real network conditions. In simulation mode, each tool call returns a simulated task success expectation without requiring live execution. This mode allows researchers to evaluate novel architectures or algorithms efficiently and deterministically, such as investigating the impact of tool description text on semantic matching. This design is free from external service dependencies, ensuring repeatable experimental conditions.
\end{itemize}

In the end, these capabilities make the MCP servers module a flexible and essential component for evaluating tool routing algorithms under both realistic and controlled environments.

\textbf{Module 2: Network Status Environment.}
This module generates configurable network status profiles for each MCP server, emulating various network conditions to support realistic and reproducible testing. It provides two main functions:

First, configuration files are generated to specify the network behavior of each server, either through randomized or predefined settings. An example of such configuration files is provided in Fig.~\ref{fig:hybrid_scenario}. These configurations cover a variety of typical network scenarios, including fluctuating latency, intermittent outages, high latency, high jitter, and near-ideal conditions. The detailed specifications of each scenario are:
\begin{enumerate}
    \item \textit{Fluctuating Latency}: Network performance follows a periodic oscillation pattern with recognizable rhythms.
    In real deployments, congestion often increases delay during the daytime and decreases at night, and different servers follow distinct patterns with varying amplitudes and periods.
    NetMCP reproduces this behavior using sinusoidal functions, configuring phase shifts, amplitudes, and periods to emulate heterogeneous load rhythms.
    
    \item \textit{Intermittent Outage}: Service availability drops to zero in random intervals of varying duration.
    Such failures are common in large-scale distributed systems due to server crashes, network partitions, or regional outages.
    NetMCP simulates outages with a configuration where probability, duration, and severity can be specified (e.g., latency fixed at $1000$ {\rm{ms}} during downtime).

    \item \textit{High Latency}: End-to-end communication is characterized by consistently elevated delays.
    This corresponds to long-distance connections such as transoceanic links or persistent congestion on bandwidth-limited paths, often seen in edge or legacy infrastructure.
    NetMCP models this by assigning a fixed high baseline (e.g., $350$ {\rm{ms}}) with low variance (e.g., $50$ {\rm{ms}}), while disabling failure or periodic oscillation modules.

    \item \textit{High Jitter}: Latency fluctuates unpredictably around a moderate baseline with high variance.
    This condition is typical for wireless networks (e.g., 5G/WiFi), congested public internet routes, or shared bandwidth environments where packet delivery times could be unpredictable.
    NetMCP simulates jitter by setting a baseline (e.g., $100$ {\rm{ms}}) with large standard variance (e.g., $70$ {\rm{ms}}) and injecting Gaussian noise.

    \item \textit{Ideal Conditions}: Network connections remain consistently stable.
    This represents intra-datacenter communication or high-speed metropolitan links, serving as a performance upper bound. 
    NetMCP configures minimal latency (e.g., $30$ {\rm{ms}}) with very small variance (e.g., $5$ {\rm{ms}}), with all failure and fluctuation modules disabled, providing a reference baseline for evaluation.  
\end{enumerate}

Second, the module uses configurations to generate historical latency time series data for each server, simulating network performance over extended periods such as $24$ hours. NetMCP can retrieve the latency sequence up to any specified time index during batch testing, providing simulated historical network states. This allows for incorporating perceived network conditions into scheduling decisions or other adaptive algorithms.

\textbf{Module 3: Agent.}
This module is a core component of NetMCP and is responsible for coordinating interactions among user queries, distributed MCP tools, and the LLM. It abstracts the complexities of tool invocation, multi-turn dialogue management, and response synthesis, providing a unified and robust interface for reasoning-based tool use. Its key capabilities are as follows:

First, it manages multi-party interactions. On the user side, the module leverages the LLM to interpret queries and generate JSON-formatted invocation parameters for the MCP tool selected by the routing algorithm, ensuring context-aware tool usage. On the tool side, it handles end-to-end API communication, including request construction, execution, and response parsing.

Second, the module facilitates multi-turn dialogue through a call-chat mechanism, alternating between tool calls and LLM-based dialogue within a single query session. Results returned by tool calls are evaluated in the chat phase. If the task is considered complete, no further calls are issued. Otherwise, the cycle continues until the task is fulfilled, dependency conditions are satisfied, or the maximum number of turns is reached. The module then aggregates all intermediate outputs and uses the LLM to generate a coherent final response. It also incorporates exception handling to address network delays, server timeouts, and execution errors, ensuring robustness under diverse operating conditions.

By encapsulating low-level communication logic and offering a consistent execution interface, the agent module reduces the complexity of tool routing, supporting reliable coordination of tools across diverse network environments.

\textbf{Module 4: Tool Routing Algorithm.}
This module provides a configurable algorithm interface that enables researchers to integrate custom tool routing strategies through a standard API, supporting rapid prototyping and comparative experiments. NetMCP also includes several baseline algorithms for direct use, such as semantic-only algorithms and the proposed SONAR algorithm, which jointly considers semantic relevance and network conditions. This design improves adaptability and extensibility, allowing systematic evaluation of routing performance across a wide range of application scenarios.

\textbf{Module 5: Evaluation Metrics.}
This module is responsible for evaluating the quality of agent responses. It takes as input the generated answer, the user's original query, and the ground truth from the query dataset, and applies the LLM-as-a-judge method~\cite{llmasajudge} to assign a quality score to the response. Based on these evaluations, the module computes a set of performance indicators, including selection success rate, average tool routing latency, expected expertise, average call tool latency, average select tool latency, and failure rate:

\begin{itemize}
    \item \textit{Selection Success Rate (SSR): }The ratio of tasks where a web search-enabled MCP server is correctly selected. It is calculated as the number of successful selections divided by the total number of tasks.
    
    \item \textit{Expected Expertise (EE): }The average expertise of the selected servers across all tasks. It is computed as the sum of the expertise values of the chosen servers divided by the total number of tasks.
    
    \item \textit{Average Latency (AL): }The average network latency (in milliseconds) of the selected servers for all web search tasks.
    
    \item \textit{Select Latency (SL): }The average tool selection latency (in milliseconds) across all user queries.
    
    \item \textit{Failure Rate (FR):} The proportion of web search task executions that encounter server failure events. In our experiments, a latency exceeding 1000ms is considered a server outage (downtime).

\end{itemize}

These metrics are exposed via the metrics APIs, enabling quantitative analysis and comparison of routing strategies under diverse network conditions.

\subsection{Workflow}

We consider the NetMCP server selection to be an optimization problem executed for each incoming query from users. Relevant concepts and notations are defined in Table~\ref{tab:symbol}.

\begin{table}[t]
\centering
\caption{Symbols and their meanings in the MCP server selection problem}
\label{tab:symbol}
\begin{tabular}{@{}cl@{}}
\toprule
Symbol & Meaning \\
\midrule
$N$ & The number of available MCP servers \\
$\mathcal{M}$ & Set of $N$ available MCP servers: $\{1, 2, \dots, N\}$ \\
$m$ & A specific MCP server, where $m \in \mathcal{M}$ \\
$t_{m,j}$ & The $j$-th tool provided by server $m$ \\
$d_m$ & Textual description of server $m$ \\
$d_{m,j}$ & Textual description of tool $t_{m,j}$ \\
$\mathcal{Q}$ & Universe of all possible user queries \\
$q$ & A specific user query ($q \in \mathcal{Q}$) \\
$t^*$ & The final selected tool \\
$m^*$ & The final selected server hosting $t^*$ \\
\bottomrule
\end{tabular}
\end{table}

Our NetMCP platform operationalizes the MCP tool routing problem through a structured, multi-stage workflow. Specifically, for each incoming user query \( q \in \mathcal{Q} \), the platform executes the following sequence of steps:

\textbf{Step 1: Build MCP Server Set.}
The platform begins by constructing the set of available servers \( \mathcal{M} = \{1, 2, \dots, N\} \), each with associated textual descriptions \( d_m \) for servers and \( d_{m,j} \) for their respective tools. This set can be built from real-world servers or simulated instances, ensuring a diverse and configurable pool for experimentation.

\textbf{Step 2: Generate Network Environment.}  
Each server \( m \in \mathcal{M} \) is assigned a network profile, including dynamic attributes such as latency and downtime. These parameters can be generated in a random pattern or manually specified by the user by routing components.

\textbf{Step 3: Tool Routing.}  
The routing process consists of three sub-steps:
\begin{itemize}
    \item \textit{Tool Prediction}: The raw user query \( q \) is transformed into a standardized tool-type description using an LLM, facilitating consistent semantic interpretation.
    \item \textit{Semantic Tool Routing}:
    Functionally relevant MCP tool candidates are identified from all available tools through a coarse-to-fine retrieval process.
    This process operates in two stages: first, the transformed query is matched against server descriptions (\( d_m \)) to coarsely preselect a subset of relevant servers; subsequently, the query is matched against the tool descriptions (\(d_{m,j} \)) within these preselected servers to determine the final, refined set of candidate tools. 
    Semantic relevance can be computed using methods such as Best Matching 25 (BM25)~\cite{BM25}, cosine similarity, or other embedding-based similarity measures.
    \item \textit{Network-Aware Tool Routing}: Each server’s recent network performance is scored based on real-time latency history. A joint optimization algorithm combines semantic relevance and network scores via a weighted utility function, selecting the final tool \( t^* \) hosted by \(m^*\), 
\end{itemize}

\textbf{Step 4: Call MCP Tool.}
The agent invokes the tool \( t^* \) on server \( m^* \), handles the execution, and processes the response. Execution results, including success status and actual latency, are recorded and returned to the user.

\textbf{Step 5: Evaluate.}  
The agent’s final response is evaluated by an LLM-based judge. Key performance metrics, as described in module 5 of section~\ref{sec:Platform}, are computed and logged through the Metrics API for subsequent analysis.

The workflow is designed in a feedforward manner. Execution latency from each tool call is recorded into the corresponding server’s historical network profile, ensuring that future routing decisions reflect up-to-date performance data. 

\section{Semantic-Oriented and Network-Aware Routing Algorithm}
\label{subsec:joint_optimization}
\begin{figure*}[!t]
\centering
\includegraphics[width=\textwidth]{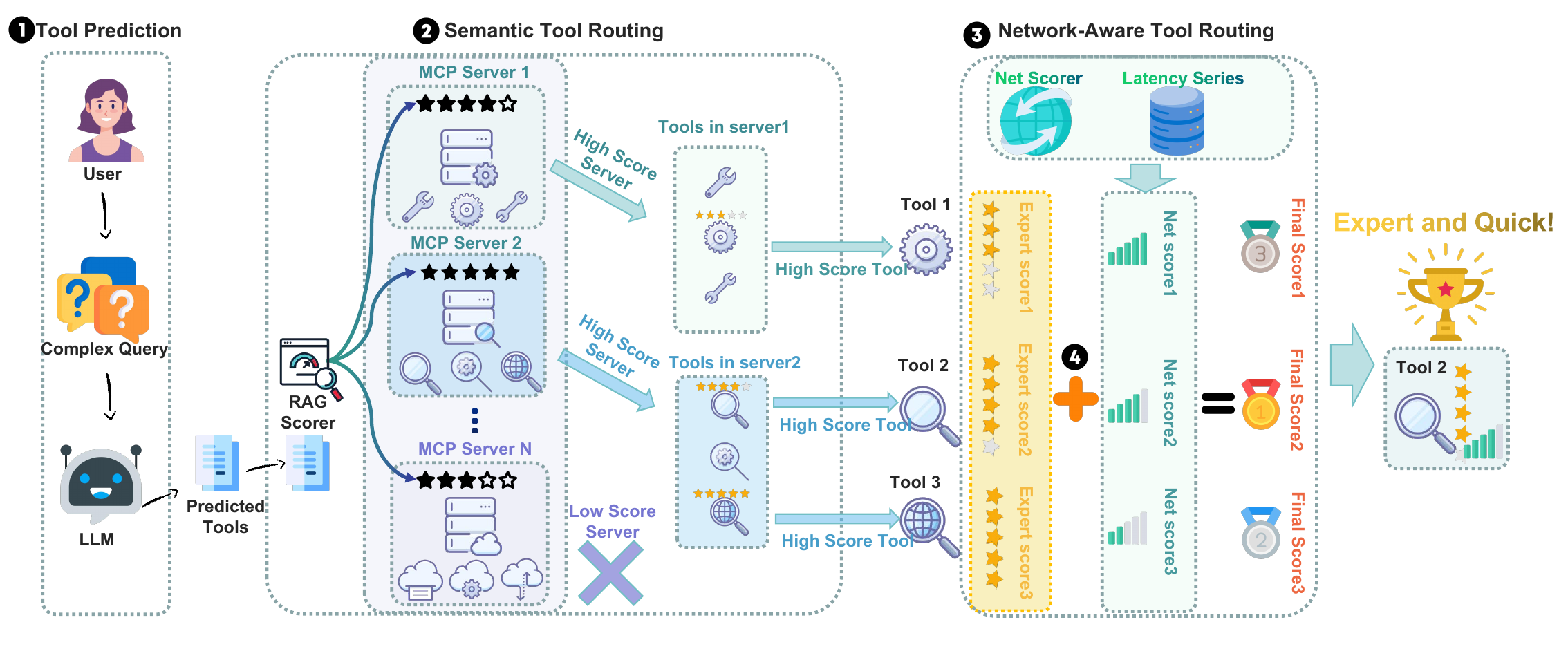}
\caption{Overview of the SONAR algorithm, which combines tool prediction, semantic-oriented tool routing, and network-aware tool routing into a joint optimization strategy for selecting the most suitable MCP tool.}
\label{fig_4}
\end{figure*}

In this section, we propose SONAR to optimize tool selection in MCP systems. It aims to identify the most suitable tool by maximizing a composite utility function that balances functional capability with network performance. The overall design of the algorithm is depicted in Fig.~\ref{fig_4}.

\begin{algorithm}[t]
\caption{SONAR Algorithm}
\label{alg:SONAR_alg}
\begin{algorithmic}[1]
\Procedure{SelectServer}{$q, \mathcal{M}, \alpha, \beta, S, T$}
    \State $q_{\text{pre}} \gets \text{LLM}_{\text{Preprocess}}(q)$
    \State $\mathcal{M}_{\text{candidate}} \gets \text{Top}_S(\{\text{BM25}(d_m, q_{\text{pre}}) \mid m \in \mathcal{M}\})$
    
    \For{$m \in \mathcal{M}_{\text{candidate}}$}
        \State $C(m) \gets \max(\{\text{BM25}(d_{m,j}, q_{\text{pre}}) \mid j \in \text{Top}_T(\mathcal{T}_m)\})$
        \State $N(m) \gets f_{\text{network}}(\mathbf{L}_m)$
        \State $S(m) \gets \alpha \cdot C(m) + \beta \cdot N(m)$
    \EndFor
    
    \State $m^* \gets \arg\max_{m \in \mathcal{M}_{\text{candidate}}} S(m)$
    \State $t^* \gets \arg\max_{j \in \mathcal{T}_{m^*}} \text{BM25}(d_{m^*,j}, q_{\text{pre}})$
    \State \Return $(m^*, t^*)$
\EndProcedure
\end{algorithmic}
\end{algorithm}

\subsection{Tool Prediction}
\label{toolprediction}
The first step of SONAR is tool prediction, which addresses key challenges in interpreting user queries through semantic abstraction. Raw inputs often contain redundant phrasing that obscures the core intent, while linguistic mismatches between queries and tool metadata (e.g., Chinese queries versus English tool descriptions) hinder direct alignment. Additionally, overly detailed queries can mislead the LLM and trigger irrelevant tool selections. 
For example, the query ``{\textit{Who founded the first luxury goods company?}}'' might incorrectly trigger a LinedIn search tool due to superficial keyword ``{\textit{company}}'' overlaps. To address these issues, SONAR applies an LLM-based preprocessing stage that transforms each raw query $q$ into a standardized tool-type description $q_{\text{pre}}$ (e.g., ``{\textit{a websearch tool}}''), ensuring reliable input for subsequent routing.


\subsection{Semantic-Oriented Tool Routing}
\label{subsec:tool_selection}

SONAR employs a two-stage coarse-to-fine retrieval method to identify the semantically most relevant MCP tools from a large candidate pool. This approach first identifies capable servers and then selects the most appropriate tools within those servers as shown in step 2 of Fig.~\ref{fig_4}.

\textbf{Server-Level Filtering.}
The preprocessed query $q_{\text{pre}}$ is matched against server descriptions using BM25:
\begin{equation}
\text{score}_{\text{server}}(m, q_{\text{pre}}) = \text{BM25}\left(d_m, q_{\text{pre}}\right),
\end{equation}
where $d_m$ represents the metadata description of server $m$. The top-$S$ servers form the candidate set as:
\begin{equation}
\mathcal{M}_{\text{candidate}} = \underset{m \in \mathcal{M}}{\text{Top}_S} \, \text{score}_{\text{server}}\left(m, q_{\text{pre}}\right).
\end{equation}
This stage reduces the search space from thousands of servers to a manageable subset while preserving high-recall coverage.

\textbf{Tool-Level Ranking.}
For candidate server in $\mathcal{M}_{\text{candidate}}$, we define the corresponding tool set as $\mathcal{T}_{\text{all}} = \bigcup_{m \in \mathcal{M}_{\text{candidate}}} \mathcal{T}_m$.
Each tool in $\mathcal{T}_{\text{all}}$ is scored based on its semantic relevance to the $q_{\text{pre}}$ as:
\begin{equation}
s_{\text{i}} = \text{BM25}\left(d_i, q_{\text{pre}}\right),
\end{equation}
where $d_{u}$ denotes the description of the tool $i$ in $\mathcal{T}_{\text{all}}$. The top-$K$ tools are selected as:
\begin{equation}
\mathcal{T}_{\text{candidate}} = \operatorname{Top}_K \left( \left\{ s_i \mid i \in \mathcal{T}_{\text{all}} \right\} \right).
\end{equation}

Moreover, since the tool expertise score derived from BM25 can range from $-\infty$ to $+\infty$, we apply softmax normalization to map these scores into a probability distribution. The softmax transformation is defined as:
\begin{equation}
C(i) = \frac{\exp(s_i)}{\sum_{j} \exp(s_j)},
\end{equation}
where $s_i$ is the BM25 score for tool $i$. This normalization amplifies the relative differences between expert tools (high $s_i$) and non-expert tools (low $s_i$), as demonstrated by the exponential scaling: when $s_i \gg s_j$, $C(i) \to 1$ while $C(j) \to 0$; when $s_i \approx s_j$, $C(i) \approx C(j)$.
This behavior enhances the separability between high-quality and low-quality tools, clarifying the ranking distinction for downstream applications.

This hierarchical design enables efficient filtering across large-scale servers and precise tool-level matching, ensuring accurate semantic-oriented selection. Yet it does not account for network conditions, which we address in the following.

\subsection{Network-Aware Tool Routing}
\label{subsec:network_aware}

We extend SONAR with a network performance scoring mechanism that incorporates real-time network states into the MCP tool routing as shown on line 6 of Algorithm~\ref{alg:SONAR_alg}. This mechanism evaluates each server’s latency characteristics and enables routing decisions to reflect current network conditions.

For each $i$ in the candidate tool set $\mathcal{T}_{\text{candidate}}$, we evaluate tool $i$'s host server $m$ with latency history $L_m = [l_1, l_2, \ldots, l_t]$ up to the current time $t$, the network score $N(i)$ is computed through a multi-dimensional assessment as:
\begin{equation}
N(i) = \text{score\_server\_latency}\left(L_m, t\right),
\end{equation}
where the scoring function incorporates four key network performance aspects:
\begin{itemize}
    \item \textit{Base score}: Base score calculated using an improved smoothing function that penalizes values beyond the ideal range (20-50 ms). It serves as the baseline performance metric upon which all subsequent penalty adjustments are proportionally applied.
    \item \textit{High Latency Penalty}: The latency predicted by Exponentially Weighted Moving-Average (EWMA)~\cite{ewma} is used for high-latency detection with proportional penalties scaled by relative excess over ideal thresholds.
    \item \textit{Trend Penalty}: Penalty for recent increasing latency trends.
    \item \textit{Outage Risk Penalty}: Penalty for recent occurrences of high latency ($> 800$ ms).
    \item \textit{Instability Penalty}: Penalty for high coefficient of variation in recent latency values.
\end{itemize}

The final network score of the tool $i$ is computed as:
\begin{align}
N(i) = & \text{base\_score} \times \left(1 - w_1 P_{\text{high}}\right) \left(1 - w_2 P_{\text{trend}}\right) \notag \\
& \times \left(1 - w_3 P_{\text{outage}}\right) \left(1 - w_4 P_{\text{instability}}\right),
\end{align}
where $\text{base\_score}$ denotes the baseline performance score ranging from 0.0 to 1.0, $w_1$ to $w_4$ represent the weight coefficients for each penalty term, and $P_{\text{high}}$, $P_{\text{trend}}$, $P_{\text{outage}}$, $P_{\text{instability}}$ denote the penalty values for high latency, trend, outage risk, and instability, respectively.

Moreover, if the current latency $l_t \geq 1000$ ms, the hosting server is treated as offline and the network score is set to $N(i) = -1$.

This network status evaluation introduces a complementary perspective by explicitly accounting for network conditions, allowing routing decisions to move beyond purely semantic relevance and incorporate the real-time quality of connections.

\subsection{Joint Optimization Strategy in SONAR}
\label{joint_method}
In SONAR, tool selection must consider semantic expertise and network performance jointly. We therefore formulate tool routing as a joint optimization problem that integrates these two dimensions into a single decision process, corresponding to the step 4 in Fig.~\ref{fig_4}.

\textbf{Selection Rule.}
We adopt a linear weighted sum to compute a composite score, combining the tool's semantic relevance to the query with the network status of its hosting MCP server. For each $i$ in the tool candidate set, 
\begin{equation}
    S(i) = \alpha \cdot C(i) + \beta \cdot N(i),
\end{equation}
where $\alpha, \beta$ are non-negative weighting coefficients that satisfy $\alpha + \beta = 1$. The parameter $\alpha$ reflects the importance assigned to semantic relevance, while $\beta$ reflects the importance assigned to network performance.
The final selection is made by choosing the tool with the highest composite score as:
\begin{equation}
i^* = \underset{i \in \mathcal{T}_{\text{candidate}}}{\arg\max} \, S(i).
\end{equation}

\textbf{Weighted Utility Maximization.}
The coefficients $\alpha$ and $\beta$ control the trade-off between semantic relevance and network performance in the MCP tool routing decision:
\begin{itemize}
    \item \textit{Quality-Priority Mode} ($\alpha > \beta$): Prefers servers with higher expertise scores, accepting longer latency when task quality is critical. This mode is suitable for applications where correctness dominates responsiveness, such as scientific data analysis, code synthesis, or legal and medical queries, where an inaccurate answer could cause serious consequences.
    \item \textit{Latency-Sensitive Mode} ($\alpha < \beta$): Prioritizes servers with lower latency, suitable for real-time interactive applications such as dialogue systems, live tutoring, or autonomous agents where user experience depends heavily on responsiveness.
    \item \textit{Balanced Mode} ($\alpha = \beta$): Seeks an equilibrium between semantic capability and network responsiveness. This mode fits general-purpose scenarios, such as knowledge assistants or productivity applications, where accuracy and timely response are essential but neither dominates.
\end{itemize}
SONAR improves overall utility by ensuring that both semantic relevance and network conditions are considered. 

\section{Experiment}
\label{experiment}
In this section, we conduct extensive experiments on the NetMCP platform to evaluate the effectiveness of our proposed network-aware algorithm, SONAR.
\subsection{Experiment Setup}
With the aid of NetMCP, we constructed a simulation environment comprising $15$ MCP servers to evaluate tool routing algorithms, including our proposed SONAR. 
Among these servers, five of them provide web search tool invocation capabilities, while the remaining ten servers provide tools for unrelated tasks such as code modification or product search on Amazon. 

To ensure consistent functionality and avoid confounding effects from mismatches between descriptions and actual MCP tool capabilities, five web search servers share the same backend implementation, based on the Exa MCP server. Their textual descriptions, however, were diversified by polishing and rephrasing with an LLM (i.e., Qwen 3-32B). This generates heterogeneous server descriptions while preserving identical underlying functionalities. As a result, observed variation in server selection can be attributed solely to semantic matching or network conditions, rather than differences in tool capability.

To evaluate our approach, we utilize MCPBench~\cite{mcpbench}, a query dataset for web search tasks. The evaluation is conducted using the metrics introduced in Section~\ref{sec:Platform}, considering both task completion accuracy and efficiency. 







\begin{figure*}[!t]
\centering
\includegraphics[width=1.9\columnwidth]{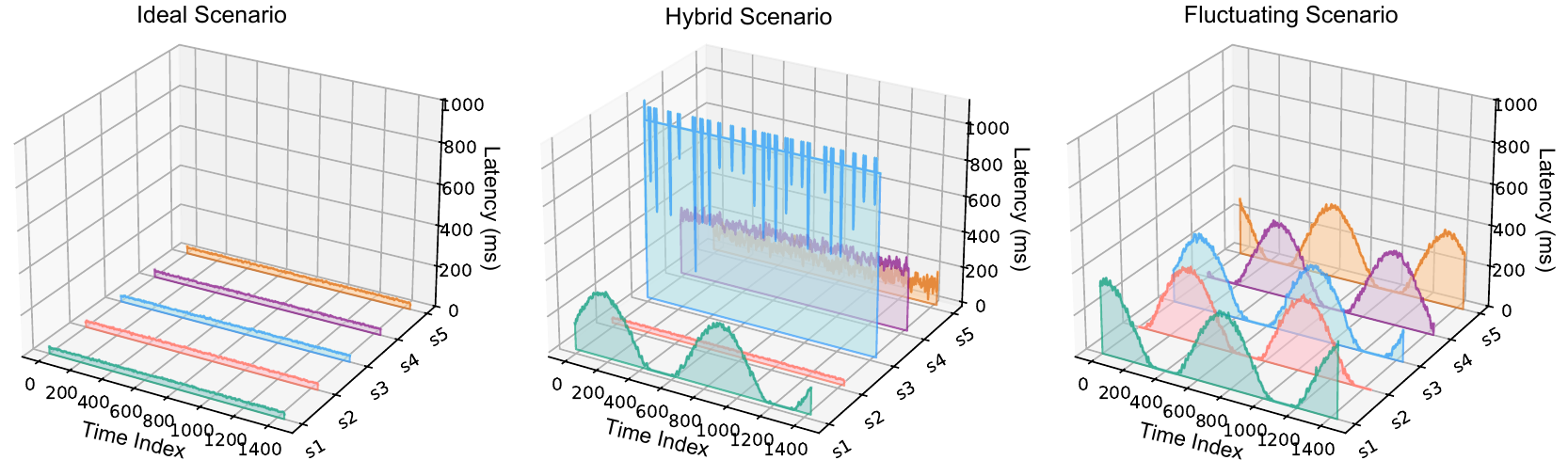}
\caption{Three network scenario settings. In the Ideal Scenario, all servers operate with stable low latency. In the Hybrid Scenario, servers exhibit heterogeneous network conditions (e.g., downtime, high jitter, high latency). In the Fluctuating Scenario, server latency varies sinusoidally over time with distinct phase offsets. }
\label{fig:Scene}
\end{figure*}

\subsection{Results and Analysis}
\label{sec:exp}
To comprehensively evaluate different MCP tool routing algorithms, we design experiments under three simulated network environments, i.e., ideal, hybrid, and fluctuating, as well as under real-world MCP tool conditions. 
As shown in Fig.~\ref{fig:Scene}, we visualize the latency curves in each scenario, where network conditions range from consistently low delay to heterogeneous disruptions and periodic fluctuations. These scenarios reflect representative patterns of network dynamics observed in practice. 
By separating experiments across different settings, we aim to examine how different routing algorithms behave when the underlying network conditions vary in stability and complexity. Specifically, we evaluate SONAR alongside three baseline tool routing algorithms, as detailed below:
\begin{enumerate}
\item \textit{RAG}: We first translate the user query to ensure it is in English. Then, following the two-stage semantic retrieval method described in Section~\ref{subsec:tool_selection}, we select the most relevant MCP tool based on BM25 similarity. This approach corresponds to the tool retrieval method implemented in the MCP Zero~\cite{MCPZero}.

\item \textit{RerankRAG}: Building upon the RAG method, we introduce an LLM-based reranking step over the candidate tool set. This allows us to refine the selection and identify the MCP tool that is most semantically aligned with the user’s query.

\item \textit{Prediction-Enhanced RAG (PRAG)}: We first apply a tool prediction step based on RAG, as detailed in Section~\ref{toolprediction}, to mitigate the impact of redundant or noisy information in the raw user query. Subsequently, using the processed query output from the tool prediction stage, we perform base RAG semantic retrieval to determine the final tool.

\end{enumerate}

\begin{figure}[!t]
\centering
\includegraphics[width=0.9\columnwidth]{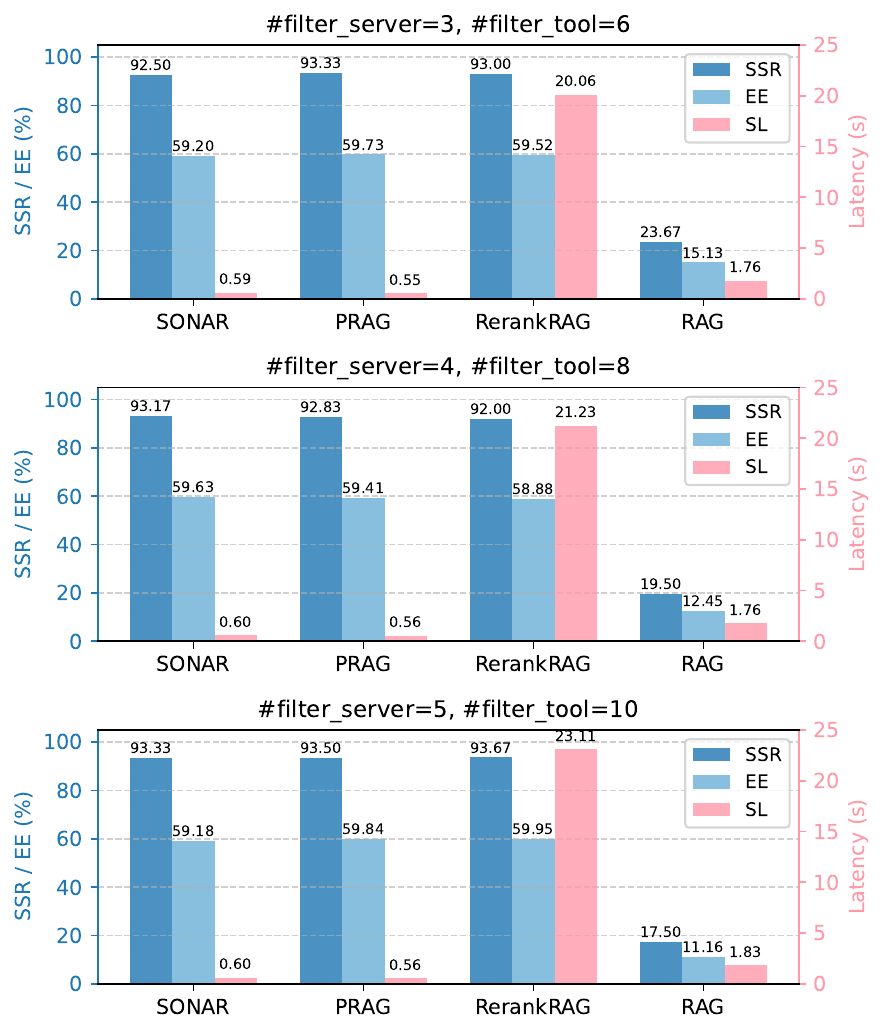}
\caption{Comparison of four tool routing algorithms, i.e., SONAR, PRAG, RerandRAG, and RAG, under ideal network conditions, measured by SSR, EE, and SL.}
\label{4methods_3metrics}
\end{figure}

\textbf{Ideal Scenario.}
In the ideal scenario, all MCP servers operate smoothly with consistently low latency, highlighting the intrinsic effectiveness of different tool routing algorithms when network conditions pose no limitations.
Fig.~\ref{4methods_3metrics} compares four tool routing algorithms under ideal network conditions using three metrics, i.e., SSR, EE, and SL. 
The baseline RAG method achieves a web search tool selection accuracy of only approximately $20\%$ because it performs no query preprocessing. 
In contrast, all other algorithms, which incorporate the tool prediction module, reach success rates of up to $90\%$. However, RerankRAG introduces substantial overhead, as its LLM-based reranking step leads to an average tool selection time exceeding $20$ seconds per query, significantly increasing end-to-end delay. In comparison, PRAG and its network-aware extension, i.e., SONAR, achieve high SSR and EE while keeping tool selection latency consistently low, benefiting from their tool prediction module, making them more suitable for tool routing.

\textbf{Hybrid Scenario.}
\label{para:Hybrid Scenario}
A hybrid network scenario is shown in the central panel of Fig.~\ref{fig:Scene}. In this setting, we configure $10$ {\textit{non-websearch}} expert MCP servers to operate under low-latency conditions, with an average response latency of 30 ms. 
In contrast, the 5 {\textit{websearch}}-capable MCP servers are assigned five distinct latency profiles to simulate diverse real-world network behaviors: fluctuating latency, downtime outage, high latency, low latency but high jitter, and ideal low latency. This setup enables comprehensive evaluation of system resilience and routing intelligence under heterogeneous and realistic network conditions.
\begin{table}[t]
  \centering
  \caption{Performance comparison between PRAG and SONAR under the Hybrid Scenario across multiple metrics, including SSR, EE, AL, and NF when semantic and network factors are assigned equal weight.}
  \resizebox{\columnwidth}{!}{
    \begin{tabular}{c|c|ccccc}
    \toprule
    Configuration & Method &  SSR (\%) &  EE (\%) & AL (ms) & FR (\%) \\
    
    \midrule
    \multicolumn{1}{c|}{\multirow{2}[2]{*}{\shortstack{\#filter\_server=3\\\#filter\_tool=6}}} & PRAG & 94.2 & 59.4 & 909.30 & 96.3 \\
          & SONAR &  93.5 & 57.0 & 22.36 & 0 \\
          
    \midrule
    \multicolumn{1}{c|}{\multirow{2}[2]{*}{\shortstack{\#filter\_server=4\\\#filter\_tool=8}}} & PRAG &  92.8 & 58.6 & 894.74 & 96.1 \\
          & SONAR &  92.0 & 56.0 & 22.822 & 0 \\
          
    \midrule
    \multicolumn{1}{c|}{\multirow{2}[2]{*}{\shortstack{\#filter\_server=5\\\#filter\_tool=10}}} & PRAG &  91.8 & 57.9 & 888.27 & 96.4 \\
          & SONAR &  93.3 & 56.8 & 22.44 & 0 \\

    \midrule
    \multicolumn{1}{c|}{\multirow{2}[2]{*}{\shortstack{\#filter\_server=6\\\#filter\_tool=12}}} & PRAG &  98.0 & 61.8 & 896.283 & 91.0 \\
          & SONAR &  96.5 & 58.9 & 21.277 & 0 \\

    \bottomrule
    
    \end{tabular}}
  \label{tab:Hybrid_Scenario}%
\end{table}%

\begin{figure*}[!t]
\centering
\includegraphics[width=1.9\columnwidth]{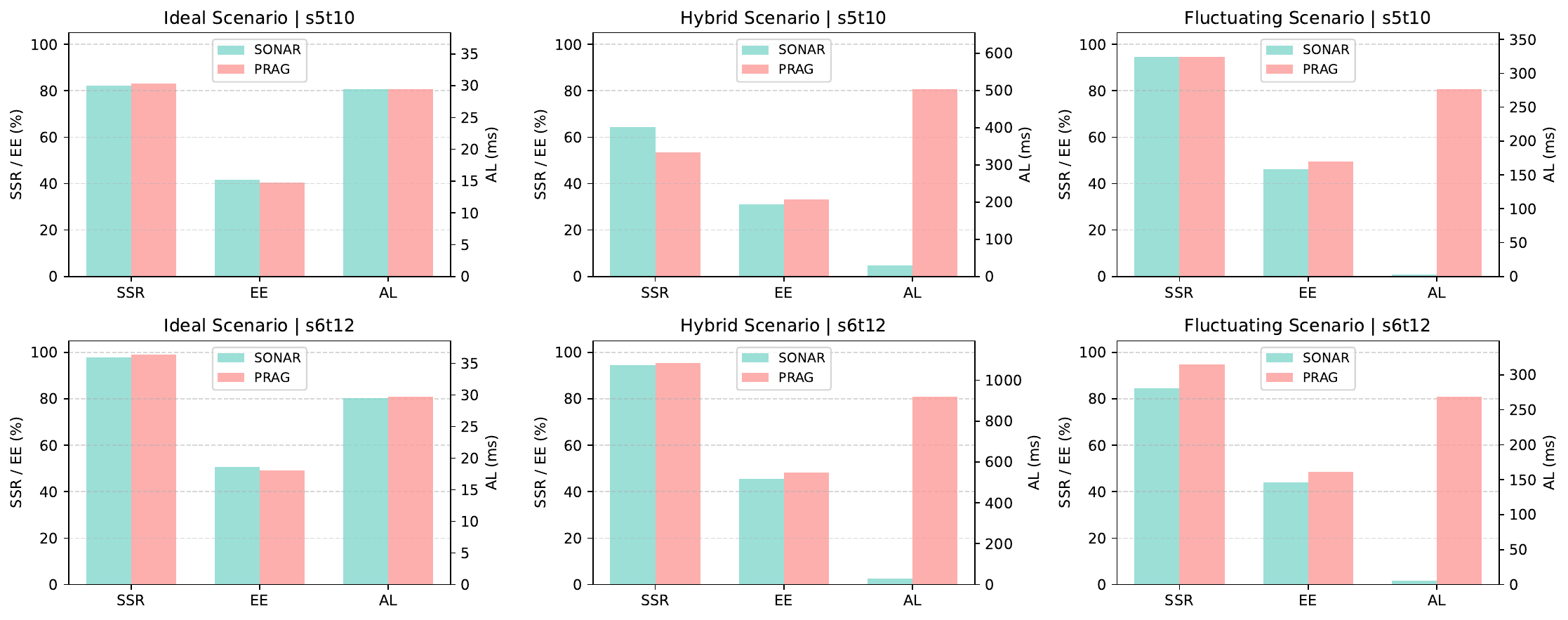}
\caption{Performance comparison between PRAG and SONAR algorithms across three network scenarios (ideal, hybrid, and fluctuating)  during real-world MCP tool invocation by the agent module, evaluating metrics including SSR, EE, and AL. }
\label{fig:real}
\end{figure*}


\begin{table}[t]
  \centering
  \caption{Performance comparison between PRAG and SONAR under the Fluctuating Scenario across multiple metrics, including SSR, EE, and AL.}
  \resizebox{\columnwidth}{!}{
    \begin{tabular}{c|c|ccc}
    \toprule
    Configuration & Method &  SSR (\%) & EE (\%) & AL (ms) \\
    
    \midrule
    \multicolumn{1}{c|}{\multirow{2}[2]{*}{\shortstack{\#filter\_server=3\\\#filter\_tool=6}}} 
    & PRAG &  0.922 & 0.582 & 161.24 \\
    & SONAR &  0.935 & 0.587 & 97.44 \\
    
    \midrule
    \multicolumn{1}{c|}{\multirow{2}[2]{*}{\shortstack{\#filter\_server=4\\\#filter\_tool=8}}} & PRAG & 92.8 & 58.6 & 162.59 \\
          & SONAR &  91.8 & 57.2 & 23.83 \\
          
    \midrule
    \multicolumn{1}{c|}{\multirow{2}[2]{*}{\shortstack{\#filter\_server=5\\\#filter\_tool=10}}} & PRAG &  93.7 & 59.1 & 160.96\\
    & SONAR &  93.2 & 58.0 & 4.41 \\

    \midrule
    \multicolumn{1}{c|}{\multirow{2}[2]{*}{\shortstack{\#filter\_server=6\\\#filter\_tool=12}}} & PRAG &  98.8 & 62.4 & 165.735\\
    & SONAR &  93.8 & 58.4 & 4.028 \\

    \bottomrule
    \end{tabular}}
  \label{tab:Fluctuating_Scenario}%
\end{table}%

Following the comparative evaluation of multiple algorithms under ideal network conditions, we select PRAG as the baseline due to its strong overall performance. 
The direct comparison between PRAG and SONAR is informative because the only difference lies in its network awareness, allowing us to isolate the effect of this capability.
Table~\ref{tab:Hybrid_Scenario} presents results under different hyperparameter settings across stages of semantic tool routing.
For example, when filtering first to five candidate servers and then to ten candidate tools, PRAG and SONAR obtain similar EE scores. However, PRAG frequently routes requests to the top-ranked tool located on a server undergoing downtime, leading to failure rates of about $90\%$. In contrast, SONAR leverages historical latency observations to evaluate both tool expertise and server network reliability jointly. This enables it to avoid selecting the top-ranked but currently unavailable server, and instead route requests to alternative servers with stable network conditions. Consequently, SONAR achieves zero failures under the same experimental setting.


\textbf{Fluctuating Scenario.}
This scenario is considered to highlight how various routing algorithms perform when all available servers are unstable, forcing them to cope with time-varying latency.
In this setting, all five {\textit{websearch}}-capable MCP servers operate under sinusoidal latency patterns, each with a distinct phase that produces different peaks and troughs. 
This design creates a complex environment to test whether SONAR and other routing algorithms can consistently balance semantic expertise with latency reduction when all candidates undergo periodic variation.
Table~\ref{tab:Fluctuating_Scenario} compares PRAG and SONAR performance across multiple metrics, including SSR, EE, and AL. We can observe that SONAR reduces average tool selection latency by $74\%$ compared with PRAG, while achieving an SSR of $93\%$ and an EE of $58\%$, nearly matching the results under the ideal scenario. These results demonstrate that SONAR can substantially reduce latency and improve task completion efficiency without sacrificing accuracy, confirming its practicality in complex network conditions.


\begin{figure*}[t]
\centering
\includegraphics[width=1.9\columnwidth]{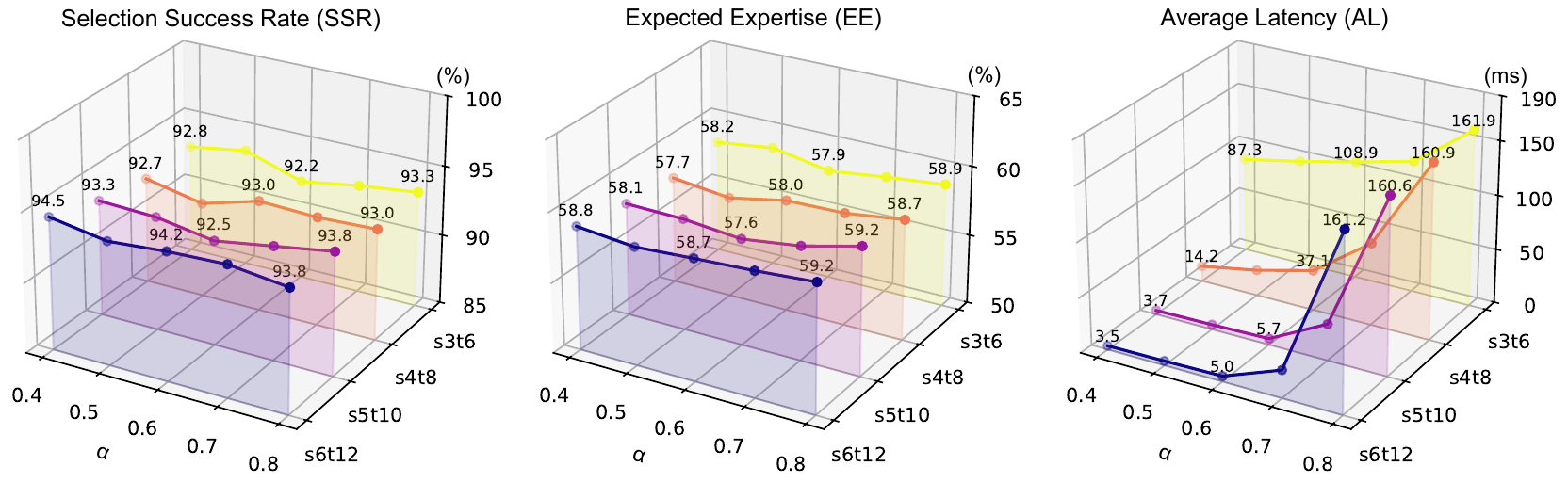}
\caption{In the Fluctuating Scenario, we compare the SSR, EE, and AL of the SONAR algorithm under different configurations.
}
\label{hyperparameter}
\end{figure*}

\textbf{Real-World Validation.}
The previous experiments validated SONAR under simulated MCP tool calls. To further confirm its effectiveness, we conducted real MCP tool invocations and compared SONAR with PRAG across the three network scenarios, as shown in Fig.~\ref{fig:real}.
In the ideal scenario, both algorithms achieved similar SSR and EE. In the hybrid scenario, PRAG frequently failed due to selecting unstable hosts. 
For example, when setting filter 6 servers and 12 tool candidates (i.e., s6t12) in NetMCP, PRAG achieves nearly $88\%$ of requests failing. In contrast, SONAR effectively incorporated historical latency information and completely avoided such failures, maintaining a $0\%$ failure rate.
In the fluctuating scenario, the two algorithms again produced comparable SSR and EE, but PRAG's average latency rose to about 300 ms, while SONAR stayed below 20 ms. Overall, the real-world results are consistent with the simulations, i.e., SONAR achieves accuracy on par with PRAG while significantly reducing latency and eliminating failures, demonstrating clear advantages in unstable network environments.

\textbf{Sensitivity Analysis.}
We introduce a joint optimization design that enables the SONAR to select MCP tools that are both highly expert and responsive in Section~\ref{subsec:joint_optimization}. This method incorporates two hyperparameters, $alpha$ and $beta$, which control the relative weights assigned to tool expertise and network status, respectively. 
To evaluate the robustness of SONAR to these design choices, we conduct a comprehensive hyperparameter analysis, systematically varying $\alpha$ and $\beta$ across a range of values and measuring their impact on key performance metrics, including SSR, EE, and AL. Results show that latency is significantly reduced when a larger portion of the routing decision accounts for network status. For instance, under the configuration of filtering six servers and selecting twelve candidate tools (i.e., s6t12), reducing $\alpha$ from 0.8 to 0.4 lowered latency from 161.2 ms to 3.5 ms without causing any drop in task success or notable decline in expertise effectiveness. This robustness demonstrates that SONAR can adapt flexibly across different parameter settings and further indicates that algorithms relying solely on semantic matching, as in most existing MCP tool routing methods, are misaligned with the requirements of real-world deployments.

\section{Future Work}
\label{sec:future_work}

This work establishes a foundational framework for network-aware tool routing in MCP systems. Several promising directions remain for future investigation:

\begin{itemize}
\item \textit{Open-sourcing and extensibility.} We plan to release the NetMCP platform as open source to encourage community engagement and collaborative development. The platform will be extended to support more advanced LLMs such as ChatGPT and Gemini, and to cover a broader range of MCP task types, thereby improving benchmark coverage and practicality.
\item \textit{Advanced joint optimization.} We intend to enhance the SONAR algorithm by exploring advanced optimization techniques beyond weighting. For example, Reinforcement Learning (RL) and other adaptive decision-making methods can be applied to capture dynamic trade-offs between semantic relevance and network conditions based on real-time feedback.
\item \textit{Multi-agent collaboration.} NetMCP will be extended to support collaborative tool invocation among multiple LLM agents, enabling system-level performance evaluation in scenarios with interdependent tasks and diverse network constraints.
\item \textit{Large-scale real-world validation.} Deployments across geographically distributed environments could be conducted to validate the robustness and generalizability of NetMCP under more real-world network conditions.
\end{itemize}
\section{Conclusion}
\label{conclusion}

We addressed the challenge of robust MCP tool routing for LLM capability extension in dynamic network environments by introducing the NetMCP platform and the SONAR algorithm. NetMCP offers a heterogeneous benchmarking environment that emulates diverse network conditions, enabling efficient and reproducible evaluation of expanded tool routing algorithms for developers. SONAR integrates semantic relevance with real-time network QoS metrics for joint optimization, achieving adaptive and reliable tool routing. Experimental results show that SONAR significantly reduces latency and failures while maintaining high task success rates, highlighting the critical value of network-aware design for production-scale LLM systems.






\bibliographystyle{unsrt}
\bibliography{ref}

\end{document}